\documentstyle[aps,preprint,eqsecnum]{revtex}
\tightenlines
\newcommand{\be}{\begin{equation}}
\newcommand{\ee}{\end{equation}}
\newcommand{\bea}{\begin{eqnarray}}
\newcommand{\eea}{\end{eqnarray}}
\newcommand{\ben}{\begin{eqnarray*}}
\newcommand{\een}{\end{eqnarray*}}
\begin{document}
\title{Anomalous Diffusion and the First Passage Time Problem}
\author{Govindan Rangarajan \thanks{Also associated with the
Jawaharlal Nehru Centre for Advanced Scientific Research,
Bangalore, India; e-mail address: rangaraj@math.iisc.ernet.in}}
\address{Department of Mathematics and Centre for Theoretical Studies\\
Indian Institute of Science, Bangalore 560 012, India}
\author{Mingzhou Ding\thanks{E-mail address: ding@walt.ccs.fau.edu}}
\address{Center for Complex Systems and Brain Sciences \\
Florida Atlantic University, Boca Raton, FL 33431}
%\date{}
\maketitle

\begin{abstract}
We study the distribution of first passage time (FPT) in Levy type of
anomalous diffusion. Using recently formulated fractional
Fokker-Planck equation we obtain three results. (1) We derive an
explicit expression for the FPT distribution in terms of Fox or
H-functions when the diffusion has zero drift. (2) For the nonzero
drift case we obtain an analytical expression for the Laplace
transform of the FPT distribution. (3) We express the FPT
distribution in terms of a power series for the case of two
absorbing barriers. The known results for ordinary
diffusion (Brownian motion) are obtained as special cases of our
more general results.
\end{abstract}
%\vskip 15pt
\newpage

\section{Introduction}

For a stochastic process, the first passage time (FPT) is defined
as the time $T$ when the process, starting from a given point,
reaches a predetermined level for the first time, and is a random
variable \cite{risken,gardiner,tuckwell,cai}. In this paper, we
consider the distribution of FPT in the context of anomalous
diffusion. By anomalous diffusion we mean a process where the mean
square displacement of the diffusive variable $X(t)$ scales with
time as $\langle X^2(t) \rangle \sim t^{\gamma}$ with $0<\gamma<2$. If
$\gamma=1$ we get ordinary diffusion. There are different
mechanisms for generating anomalous diffusion. Our focus will be
on the continuous time random walk (CTRW) \cite{shlesinger,list}
where the waiting time is a Levy type variable obeying certain
power law distribution. We subsequently refer to the process as
the Levy type of anomalous diffusion. Recent work shows that in
the generalized diffusive limit the probability density function
of the CTRW is described by a fractional Fokker-Planck equation
(FFPE) \cite{metzler2,metzler}. Using this equation, together with
appropriate initial and boundary value condition, we derive the
exact solution for the FPT density function in terms of Fox or H
functions \cite{fox,mathai} when the anomalous diffusion has zero
drift. For the nonzero drift case we derive an expression for the
Laplace transform of the FPT density function. This Laplace
transform is then used to obtain the mean and variance of the FPT.
Finally, we derive an explicit expression for the FPT density
function for Levy type anomalous diffusion with absorbing barriers
at finite distances from the origin.

\section{Levy Type Anomalous Diffusion}

For an ordinary diffusive process, the mean square displacement
scales with $t$ as $\langle X^2(t) \rangle \sim t$ in the large $t$ limit. If
$\langle X^2(t) \rangle \sim t^{\gamma}$, we have anomalous diffusion, with $0 <
\gamma < 1$ and $1 < \gamma < 2$ corresponding to subdiffusive and
superdiffusive cases, respectively. Anomalous diffusion with drift
$X_{\mu}(t)$ is defined by
\begin{equation}
X_{\mu}(t) = \mu t + X(t).
\end{equation}
For further discussions and examples of anomalous diffusions in
physical, chemical and biological systems see
\cite{shlesinger,list}. For completeness we give in what follows a
brief introduction to CTRW and the limiting process leading to the
establishment of the FFPE.

\subsection{Details of the Process}

We start with a one dimensional continuous time random walk
described by the following Langevin equation:
\be
\frac{dX}{dt} = \sum_{i=1}^{\infty} Y_i \ \delta (t-t_i). \ee
Here the random walker starts at $x=0$ at time $t_0=0$. Subsequently,
the random walker waits at a given location $x_i$ for time
$t_i-t_{i-1}$ before taking a jump $Y_i$ which could depend on the
waiting time. The waiting time $u >0$ and the jump size $y$
($-\infty < y < \infty$) are drawn from the joint probability
density function $\phi(y,u)$. The waiting time distribution
$\psi(u)$ is given by
\begin{equation}\label{psi}
\psi(u) = \int_{-\infty}^{\infty} dy \phi(y,u).
\end{equation}
The process is non Markovian if $\psi(u)$ is a non-exponential
distribution since the probability for the next jump to occur depends
on how long the the random walker has been waiting since the
previous jump. But CTRW is non Markovian in a special way since
it does not depend on the history of the process prior to the previous
jump.

It can be shown \cite{shlesinger} that the probability
distribution $W(x,t)$ for the CTRW is related to $\phi(y,u)$ and
$\psi(u)$. This relation is simpler to write down in the
Fourier-Laplace transform space. Denoting the Fourier-Laplace
transform of $W(x,t)$ and $\phi(y,u)$ by $\tilde{W}(k,s)$ and
$\tilde{\phi}(k,s)$ respectively (where $k$ is the Fourier
transform of the space variable and $s$ the Laplace transform of
the time variable) we have \cite{shlesinger}
\begin{equation}\label{Wks}
\tilde{W}(k,s) = \frac{1}{s} \frac{1-\tilde{\psi}(s)}{1-\tilde{\phi}(k,s)}.
\end{equation}
Here $\tilde{\psi}(s)$ is the Laplace transform of $\psi(u)$.

It can be further shown that, depending on the specific form of
$\phi(y,u)$, the CTRW can produce both subdiffusive ($0 < \gamma <
1$) and superdiffusive processes ($1 < \gamma < 2$) as well as
ordinary diffusion ($\gamma=1$) \cite{shlesinger,wang}. For
example, consider
\begin{equation}\label{LWpdf}
\phi(y,u) = \frac{1}{\sqrt{2 \pi \sigma^2}} \exp[-y^2/2 \sigma^2]
\frac{(\alpha-1)/\tau}{(1+u/\tau)^{\alpha}},
\end{equation}
where $y$ and $u$ are decoupled with $y$ being a Gaussian variable
with zero mean. For $1
< \alpha < 2$, the corresponding CTRW is characterized by a
subdiffusive process with $\gamma=\alpha-1$, and for $\alpha \ge
2$, one gets ordinary diffusion with $\gamma=1$. This is
demonstrated as follows.

The waiting time distribution $\psi(u)$ is given by [cf. Eqs.
(\ref{LWpdf}) and (\ref{psi})]
\begin{equation}\label{psiu}
\psi(u) = \frac{(\alpha-1)/\tau}{(1+u/\tau)^{\alpha}}.
\end{equation}
Even though this is not a Levy stable distribution,
it belongs to the domain of attraction \cite{taqqu} of a one-sided stable
Levy law. It has infinite mean for $1 < \alpha < 2$, the range we are
interested in (it has a finite mean for $\alpha > 2$).
We call $u$ a ``Levy type of variable'' for want of a
better name.
The Laplace transform $\tilde{\psi}(s)$ of $\psi(u)$ is \cite{erdelyi2}
\begin{equation}
\tilde{\psi}(s) =  (\alpha-1) (\tau s)^{\alpha-1} \Gamma
(1-\alpha, \tau s) e^{\tau s}.
\end{equation}
The Fourier-Laplace transform $\tilde{\phi}(k,s)$ of $\phi(y,u)$
in Eq. (\ref{LWpdf}) is given by \cite{erdelyi2}
\begin{equation}\label{phiks}
\tilde{\phi}(k,s) = \exp(-\sigma^2 k^2/2) \tilde{\psi}(s).
\end{equation}

To show that the CTRW characterized by Eq. (\ref{LWpdf}) exhibits
anomalous diffusion, we need to demonstrate that
$\langle X^2(t) \rangle \sim t^{\gamma}$ for large $t$ ($0 < \gamma < 1$).
By Tauberian theorem \cite{feller}, this is equivalent
to $\langle X^2(s) \rangle \sim 1/s^{1+\gamma}$ for small $s$ ($0 < \gamma < 1$).
But we have the result
\begin{equation}\label{msd}
\langle X^2(s) \rangle = - \frac{\partial^2}{\partial k^2} \tilde{W}(k,s) \vline_{k=0}.
\end{equation}
Thus we need the expression for $W(k,s)$ in the limit $s \to 0$.
Since $W(k,s)$ is a function of $\tilde{\psi}(s)$ and $\tilde{\phi}(k,s)$
[cf. Eq. (\ref{Wks})], we first consider $\tilde{\psi}(s)$.
We have \cite{gradshteyn}
\be
\Gamma(1-\alpha,\tau s) = \Gamma (1-\alpha) - \sum_{n=0}^{\infty}
\frac{ (-1)^n (\tau s)^{1-\alpha+n}}{n! (1-\alpha+n)}.
\ee
As $s \to 0$,
\be
\Gamma(1-\alpha,\tau s)  \approx  -\frac{\Gamma (2-\alpha)}{\alpha-1} +
\frac{(\tau s)^{1-\alpha}}{\alpha-1} + \frac{(\tau s)^{2-\alpha}}{2-\alpha}.
\ee
Therefore $\tilde{\psi}(s)$ as $s \to 0$ is given by
\bea \label{psis}
\tilde{\psi}(s) & \approx & 1-\Gamma(2-\alpha) (\tau s)^{\alpha-1}, \ \ \ 1 < \alpha < 2,\\
                & \approx & 1-(2 \alpha -3) \tau s/(\alpha-2), \ \ \ \alpha > 2.
\eea
Substituting this in the Eq. (\ref{Wks}) we have [cf. Eq. (\ref{phiks})]
\bea
\tilde{W}(k,s) & \approx & \frac{1}{s}
\frac{\Gamma(2-\alpha)(\tau s)^{\alpha-1}}{1-[1-\Gamma(2-\alpha) (\tau s)^{\alpha-1}]\exp(-\sigma^2 k^2/2)},
\ \ \ 1 < \alpha < 2, \\
& \approx & \frac{1}{s}
\frac{\alpha \tau s/(\alpha-2)}{1-[1-\alpha \tau s/(\alpha-2)]\exp(-\sigma^2 k^2/2)},
\ \ \ \alpha > 2.
\eea
Using this in Eq. (\ref{msd}) and evaluating the second derivative at $k=0$ we obtain
in the limit $s \to 0$
\bea
\langle X^2(s) \rangle & \sim & \frac{\sigma^2}{\Gamma(2-\alpha) \tau^{\alpha-1}} \frac{1}{s^{\alpha}},
\ \ \ 1 < \alpha < 2, \\
& \sim & \frac{(\alpha-2) \sigma^2}{\alpha \tau} \frac{1}{s^2},
\ \ \ \alpha > 2.
\eea
Thus we have shown that the CTRW characterized by
Eq. (\ref{LWpdf}) is a subdiffusive process for $1 < \alpha < 2$
with $\gamma=\alpha-1$, and for $\alpha > 2$, one gets ordinary
diffusion with $\gamma=1$. For $\alpha=2$, by taking proper
limits, one can show that we again obtain ordinary diffusion.

We verify the above result numerically by simulating the CTRW
process. For the sake of numerical efficiency, we
replace the waiting time distribution $\psi(u)$ in Eq. (\ref{psiu})
by the Pareto distribution \cite{evans}:
\bea
\psi(u) & = & 0, \ \ \ u < \tau, \\
& = & \frac{(\alpha-1) \tau^{\alpha-1}}{u^{\alpha}}, \ \ \ u \ge \tau,
\eea
This approximation is well justified for small values of $\tau$. Numerically
obtained mean square displacement is compared with the theoretical
prediction in Figure 1 for $\alpha = 1.5$ (i.e. $\gamma = 0.5$), $a=1.0$,
$\tau=10^{-4}$ and $\sigma^2 = 3.5 \times 10^{-3}$. We
note that the numerical simulation is in excellent agreement with
the theoretical prediction.

If, on the other hand, $\phi(y,u)$ is given by the coupled
distribution \cite{shlesinger}
\begin{equation}\label{LWpdf2}
\phi(y,u) = \frac{1}{2} \delta(u/\tau-|y|/\sigma)
\frac{(\beta-1)/\tau}{(1+u/\tau)^{\beta}},
\end{equation}
where $2 < \beta < 3$ and $\delta(\cdot)$ is the Dirac delta
function, the CTRW describes a superdiffusive process with
$\gamma=\beta-1$. This can be seen as follows.

Proceeding as before, the waiting time distribution $\psi(u)$
is given by [cf. Eqs. (\ref{LWpdf2}) and (\ref{psi})]
\begin{equation}\label{psiu2}
\psi(u) = \frac{(\beta-1)/\tau}{(1+u/\tau)^{\beta}}.
\end{equation}
It has a finite mean but infinite variance for $2 < \beta < 3$.
Its Laplace transform in the limit $s \to 0$ is given by
\begin{equation}
\tilde{\psi}(s) \approx 1-(2 \beta-3) \tau s/(\beta-2)
\end{equation}
The Fourier-Laplace transform $\tilde{\phi}(k,s)$ of $\phi(y,u)$
in Eq. (\ref{LWpdf2}) is given by (in the limit $s \to 0$ and
$k \ll s$)
\begin{equation}\label{phiks2}
\tilde{\phi}(k,s) \approx \tilde{\psi}(s) - \frac{(\beta-1)\Gamma (3-\beta)}{2}
k^2 \sigma^2 (\tau s)^{\beta-3}.
\end{equation}
Substituting $\tilde{\psi}(s)$ and $\tilde{\phi}(k,s)$ in Eq. (\ref{Wks}) we get
\be
\tilde{W}(k,s) \approx \left[s+\frac{(\beta-1)(\beta-2)\Gamma(3-\beta)}{2 (2 \beta-3)}
k^2 \sigma^2 (\tau s)^{\beta-3} \right]^{-1}.
\ee
Using this in Eq. (\ref{msd}) and evaluating the second derivative at $k=0$ we obtain
in the limit $s \to 0$
\be
\langle X^2(s) \rangle \approx \frac{(\beta-1)(\beta-2)\Gamma(3-\beta)}{(2 \beta-3)}
\frac{1}{s^{5-\beta}}.
\ee
By Tauberian theorem \cite{feller}, this is equivalent to
$\langle X^2(t) \rangle \sim t^{4-\beta}$.
Thus the CTRW characterized by Eq. (\ref{LWpdf2}) is a superdiffusive process
for $2 < \beta < 3$ with $\gamma=4-\beta$ ($1 < \gamma < 2$).

Note that there are several other possible forms for $\phi(y,u)$
which also give rise to sub-  and superdiffusive behavior. See
\cite{shlesinger} for a detailed discussion.

\subsection{Fractional Fokker-Planck Equation for Levy type anomalous diffusion}

In the previous subsection, we had described CTRW processes that
give rise to Levy type anomalous diffusion. However, it is
difficult to derive analytical results directly from the process.
It is more convenient to work in the general framework of
Fokker-Planck equations \cite{risken}. One can go from the CTRW
process to a fractional Fokker-Planck equation (FFPE)
\cite{metzler} by taking the generalized diffusion limit. This
limit is analogous to the regular diffusion limit that one
considers to derive the regular Fokker-Planck equation
\cite{risken} from a random walk with $\langle X^2(t) \rangle \sim t$ for
large $t$. In the regular diffusion limit, one lets $\sigma, \tau
\to 0$ such that $\sigma^2/\tau$ is maintained a constant. For a
CTRW where $\langle X^2(t) \rangle \sim t^{\gamma}$ ($0 < \gamma < 2$), the
generalized diffusion limit is obtained by taking the limit
$\sigma, \tau \to 0$ such that $\sigma^2/\tau^{\gamma}$ is
maintained a constant.

Thus to obtain a FFPE from the CTRW process, we need to take the limit
$\sigma, \tau \to 0$. We will take this limit for the Fourier-Laplace
transform $\tilde{W}(k,s)$ of $W(x,t)$ and then invert the transform
to obtain the FFPE.
First consider the subdiffusive CTRW characterized by Eq. (\ref{LWpdf}).
In the previous section, we have already derived an expression for
$\tilde{W}(k,s)$ in the limit $s \to 0$ (which is equivalent to the
limit $\tau \to 0$ that we now require):
\be
\tilde{W}(k,s) \approx \frac{1}{s}
\frac{\Gamma(1-\gamma)(\tau s)^{\gamma}}{1-[1-\Gamma(1-\gamma) (\tau s)^{\gamma}]\exp(-\sigma^2 k^2/2)}.
\ee
Here we have used $\gamma=\alpha-1$ ($0 < \gamma < 1$) instead of $\alpha$ since
it is the physically relevant quantity. Now we take the
further limit $\sigma \to 0$ such that $\sigma^2/2 \Gamma(1-\gamma) \tau^{\gamma} =
K$ is a constant. $K$ is called the generalized diffusion constant. We then obtain
\be
\tilde{W}(k,s) = \frac{1}{s + K k^2 s^{1-\gamma}}.
\ee
This can be rewritten as
\be \label{FFPEinv}
\tilde{W}(k,s) - \frac{1}{s} = - K k^2 s^{-\gamma} \tilde{W}(k,s).
\ee
To take the inverse Fourier-Laplace transform of the above equation,
we need the inverse Laplace transform of $s^{-\gamma} \tilde{W}(k,s)$.
This is given by the Riemann-Liouville fractional integral
$_0 D_t^{-\gamma} W(k,t)$ which is defined as \cite{oldham,miller}
\begin{equation}
_0 D_t^{-\gamma} W(k,t) = \frac{1}{\Gamma(\gamma)} \int_0^t \ dt'
\ (t-t')^{\gamma-1}W(k,t'), \ \ \ \gamma > 0.
\end{equation}
This result enables us to take the inverse Fourier-Laplace
transform of Eq. (\ref{FFPEinv}) giving \cite{metzler2,metzler}
\be \label{FFPEsub}
W(x,t)-W(x,0) =\  K \ _0 D_t^{-\gamma}
\frac{\partial^2}{\partial x^2} W(x,t), \ \ \ 0 < \gamma < 1. \ee
Here we have incorporated the initial condition
$W(x,0)=\delta(x)$.

Next we consider the superdiffusive CTRW process characterized by
Eq. (\ref{LWpdf2}). We have already derived an expression for
$\tilde{W}(k,s)$ in the limit $k,s \to 0$ (which is equivalent to
the limit $\sigma, \tau \to 0$ that we now require):
\be
\tilde{W}(k,s) \approx \left[s+\frac{(3-\gamma)(2-\gamma)\Gamma(\gamma -1)}{2 (5-2\gamma)}
k^2 \sigma^2 (\tau s)^{1-\gamma} \right]^{-1}.
\ee
This expression is valid in the regime $\sigma \ll \tau$. Taking the
generalized diffusion limit we again get Eq. (\ref{FFPEinv}) but
now $K =(3-\gamma)(2-\gamma)\Gamma(\gamma -1)\sigma^2/
2 (5-2\gamma) \tau^{\gamma}$ and $1 < \gamma < 2$. Therefore, this
superdiffusive CTRW process also gives Eq. (\ref{FFPEsub}) with
the above $K$ and $1 < \gamma < 2$.

To summarize, a class of Levy type anomalous diffusion can be described
by the following FFPE in the generalized diffusion limit:
\be \label{FFPE0}
W(x,t)-W(x,0) =\  K \ _0 D_t^{-\gamma} \frac{\partial^2}{\partial x^2} W(x,t),
\ \ \ 0 < \gamma < 2.
\ee
Here $K$ has different expressions for $ 0 < \gamma < 1$ and $ 1 < \gamma < 2$
as described above. We note that there are other CTRW processes which may
give rise to a different FFPE. In this paper, we consider only the FFPE
given in Eq. (\ref{FFPE0}). The above FFPE describes Levy type anomalous
diffusion with zero drift. For anomalous diffusion with drift $\mu$, the
above FFPE can be generalized to give \cite{metzler2}:
\begin{equation}
W(x,t)-W(x,0) =\  K \ _0 D_t^{-\gamma} \frac{\partial^2}{\partial x^2} W(x,t)
- \mu \  _0 D_t^{-1} \frac{\partial}{\partial x}
W(x,t), \label{FFPE2}
\end{equation}
where $0 < \gamma < 2$. We also note that for $\gamma=1$, the above FFPE
reduces to the regular Fokker-Planck equation describing Brownian motion.

We now show that this FFPE gives the correct moments. The natural
boundary conditions for the FFPE are:
\be
W(-\infty,t) = W(\infty,t) = 0, \ \ \ W(x,0) = \delta(x).
\ee
First we compute $\langle X_{\mu} \rangle$. Multiplying the FFPE given in Eq. (\ref{FFPE2})
by $x$ and integrating from $x=-\infty$ to $x=\infty$ we get
\be
\langle X_{\mu} \rangle = \mu\  _0 D_t^{-1} \int_{-\infty}^{\infty} dx\, W(x,t) - K\
_0 D_t^{-\gamma} \int_{-\infty}^{\infty} dx\, \frac{\partial W(x,t)}{\partial x},
\ee
where we have performed integration by parts and assumed that
$x W(x,t) = 0$ and $x \frac{\partial W(x,t)}{\partial x} = 0$
at $x= \pm \infty$. Evaluating the remaining integrals using the
natural boundary conditions and the normalization integral
$ \int_{-\infty}^{\infty} dx\, W(x,t)=1$, we get
\be
\langle X_{\mu} \rangle = \mu\  _0 D_t^{-1} (1) = \mu t.
\ee
This is the expected result.

Next we compute Var[$X_{\mu}$]. Multiplying the FFPE in Eq. (\ref{FFPE2}) by
$x^2$ and integrating we get
\be
\langle X_{\mu}^2 \rangle = 2 \mu \  _0 D_t^{-1} \int_{-\infty}^{\infty} dx\, x W(x,t) - 2 K\
_0 D_t^{-\gamma} \int_{-\infty}^{\infty} dx\, x \frac{\partial W(x,t)}{\partial x},
\ee
where we have again used integration by parts and assumed that
$x^2 W(x,t) = 0$ and $x^2 \frac{\partial W(x,t)}{\partial x} = 0$
at $x= \pm \infty$. Evaluating the remaining integrals using the
fact that $ \int_{-\infty}^{\infty} dx\, x W(x,t)= \langle X_{\mu} \rangle = \mu t$
and the boundary and normalization conditions we obtain
\be
\langle X_{\mu}^2 \rangle = 2 \mu^2 \  _0 D_t^{-1} (t) + 2 K\ _0 D_t^{-\gamma} (1).
\ee
But $ _0 D_t^{-1} (t) = t^2/2$ and $ _0 D_t^{-\gamma} (1) = t^{\gamma}/
\Gamma(\gamma+1)$ \cite{miller}. Therefore
\be
\langle X_{\mu}^2 \rangle = \mu^2 t^2 + \frac{2K t^{\gamma}}{\Gamma(\gamma+1)}.
\ee
The variance Var[$X_{\mu}$] is given by $\langle X_{\mu}^2 \rangle-\langle X_{\mu} \rangle^2$. Hence
\be
{\rm Var}[X_{\mu}] = \frac{2K t^{\gamma}}{\Gamma(\gamma+1)}.
\ee
Thus we see that our FFPE gives the correct moments.

\section{First passage time problem for Levy type anomalous diffusion}

We now formulate the first passage time problem for the FFPE
given in Eq. (\ref{FFPE2}) describing a Levy type anomalous
diffusion with drift.
Consider a stochastic process $X(t)$ with $X(0)=0$. The first
passage time (FPT) $T$ to the point $X=a$ is defined as
\cite{grimmet}
\begin{equation}
T = \inf \{ t: X(t) = a \}.
\end{equation}
We would like to obtain the probability density function
for $T$. This is the first passage time problem.

For a process described by Fokker-Planck equations, the problem of
obtaining the FPT density function can be recast as a boundary
value problem with absorbing boundaries \cite{risken}. In our
case, to obtain the FPT density function, we first need to solve
Eq. (\ref{FFPE2}) with absorbing boundaries at $x=-\infty$ and
$x=a$, where $a$ is the predetermined level of crossing, with the
initial condition $W(x,0)=\delta(x)$ \cite{risken}. An equivalent
formulation, due to symmetry, is to solve Eq.~(\ref{FFPE2}) with
the following boundary and initial conditions: \be \label{bc2}
W(0,t)=0,\ \ \ W(\infty,t) = 0,\ \ \ W(x,0) =\delta(x-a), \ee
where $x=a$ is the new starting point of the CTRW, containing the
initial concentration of the distribution. The equivalence is
easily seen by making the change of variables $x \to a-x$ in Eq.
(\ref{FFPE2}). The only change is in the interpretation of $\mu$.
Now $\mu < 0$ corresponds to drift towards the barrier. This
latter formulation makes the subsequent derivation less
cumbersome. Once we solve for $W(x,t)$, the first passage time
density $f(t)$ is given by \cite{risken}
\begin{equation}
f(t) = - \frac{d}{dt} \int_0^{\infty}\ dx \ W(x,t).
\label{fT}\end{equation}
In the following subsections, we obtain the FPT density function
for Levy type anomalous diffusion under different
conditions.

\subsection{Zero Drift Case}

In this section, we obtain an explicit solution
for the FPT density for a zero drift Levy type anomalous diffusion in
terms of H-functions \cite{fox,mathai,srivastava}.
Asymptotic expressions for the FPT density in various limits are
obtained. Numerical simulations are performed to confirm the
above results.

Setting $\mu=0$ in Eq. (\ref{FFPE2}) we obtain
\be \label{FFPE}
W(x,t)-W(x,0) =\  K \ _0 D_t^{-\gamma} \frac{\partial^2}{\partial x^2} W(x,t).
\ee
Taking into account of the boundary and initial conditions [cf. Eq. (\ref{bc2})]
we are led to the following expansion for $W(x,t)$ \cite{morse}
\begin{equation}
W(x,t) = \frac{2}{\pi} \int_0^{\infty}\  dk \sin kx \ \sin ka \
A(k,t), \label{sol1}\end{equation}
with $A(k,0)=1$. To determine
the unknown function $A(k,t)$, we substitute the above expansion
for $W(x,t)$ in Eq.~(\ref{FFPE}) and, after straightforward
algebra, obtain $A(k,t)-1=-K k^2\ _0 D_t^{-\gamma} A(k,t)$. Taking
the Laplace transform with respect to $t$, we have
\begin{equation}
A(k,s) = \frac{1}{s+k^2 K s^{1-\gamma}}, \label{lap}
\end{equation}
where $A(k,s)$ is the Laplace transform of $A(k,t)$. Here we have
applied the result \cite{miller} that the Laplace transform of $_0
D_t^{-\gamma} A(k,t)$ is $A(k,s)/s^{\gamma}$.

Inverse Laplace transform of Eq.~(\ref{lap}) yields \cite{erdelyi}
\begin{equation}
A(k,t) = E_{\gamma}(-k^2 K t^{\gamma}), \label{akt}
\end{equation}
where $E_{\gamma}(z)$ is the Mittag-Leffler function
\cite{erdelyi}. The Mittag-Leffler function can be defined by
the following power series expansion:
\be \label{mf}
E_{\gamma}(z) = \sum_{n=0}^{\infty} \frac{z^n}{\Gamma(n \gamma+1)}.
\ee
When $\gamma=1$, the Mittag-Leffler function reduces to the usual
exponential function $e^z$.
Substituting Eq.~(\ref{akt}) into Eq. (\ref{sol1})
we get
\begin{equation}
W(x,t) = \frac{2}{\pi} \int_0^{\infty}\  dk \sin kx \ \sin ka
\ E_{\gamma}(-k^2 K t^{\gamma}).
\label{sol2}\end{equation}

To proceed further, we introduce the Fox or H-function
\cite{fox,mathai,srivastava} which has the following alternating power series
expansion:
\begin{eqnarray}
H_{p,q}^{m,n}\left( z\; \vline
\begin{array}{c}
(a_j,A_j)_{j=1, \ldots ,p}\\
(b_j,B_j)_{j=1, \ldots ,q}
\end{array}
\right) & = & \sum_{l=1}^{m} \ \sum_{k=0}^{\infty} \ \frac{(-1)^k
z^{s_{lk}}}{k! B_l} \nonumber \\
 & & \times \frac{\prod_{j=1,j \neq l}^m \Gamma
 (b_j-B_j s_{lk}) \prod_{r=1}^n \Gamma(1-a_r+A_r
s_{lk})}{\prod_{u=m+1}^q \Gamma(1-b_u+B_u s_{lk}) \prod_{v=n+1}^p
\Gamma(a_v-A_v s_{lk})}, \label{hfn}
\end{eqnarray}
where $s_{lk} = (b_l+k)/B_l$ and an empty product is interpreted
as unity. Further, $m,n,p,q$ are nonnegative integers such that $0
\leq n \leq p$, $1 \leq m \leq q$; $A_j$, $B_j$ are positive
numbers; $a_j$, $b_j$ can be complex numbers. We will denote the
H-function by $H_{p,q}^{m,n}(z)$ for the sake of notational
simplicity wherever this does not lead to any confusion. The
H-function has several remarkable properties, some of which are
listed in the Appendix.

Returning to our problem, by comparing the series expansion
[cf. Eq. (\ref{mf})] of the Mittag-Leffler function $E_{\gamma}(z)$ with
that of the H-function [cf. Eq. (\ref{hfn})], we see that
\begin{equation}\label{mfhfn}
E_{\gamma} (-z) = H_{1,2}^{1,1}\left( z \; \vline
\begin{array}{cc}
(0,1) & \\
(0,1), & (0,\gamma)
\end{array}
\right).
\end{equation}
Substituting this in Eq. (\ref{sol2}) we obtain
\begin{equation}
W(x,t) = \frac{2}{\pi} \int_0^{\infty}\  dk \sin kx \ \sin ka
\ H_{1,2}^{1,1}\left( k^2 K t^{\gamma} \; \vline
\begin{array}{cc}
(0,1) & \\
(0,1), & (0,\gamma)
\end{array}
\right).
\end{equation}
Letting $k' = k (Kt^{\gamma})^{1/2}$ the above equation becomes
\begin{equation}
W(x,t) = \frac{2}{\pi (Kt^{\gamma})^{1/2}} \int_0^{\infty}\ dk'
\sin k'x \ \sin k'a \ H_{1,2}^{1,1}\left( (k')^2 \; \vline
\begin{array}{cc}
(0,1) & \\
(0,1), & (0,\gamma)
\end{array}
\right).
\end{equation}
Using Property 5 [Eq. (\ref{p5})] of the H-functions to replace
$(k')^2$ by $k'$ we get
\begin{equation}
W(x,t) = \frac{1}{\pi (Kt^{\gamma})^{1/2}} \int_0^{\infty}\ dk'
\sin k'x \ \sin k'a \ H_{1,2}^{1,1}\left( k' \; \vline
\begin{array}{cc}
(0,1/2) & \\
(0,1/2), & (0,\gamma/2)
\end{array}
\right).
\end{equation}

The above equation can be rewritten as follows making use of the
standard trigonometric identity $2 \sin k'x \sin k'a = \cos
k'(x-a) - \cos k'(x+a)$:
\begin{equation}
W(x,t) = \frac{1}{2 \pi (Kt^{\gamma})^{1/2}} \int_0^{\infty}\ dk'
[\cos k'(x-a) - \cos k'(x+a)] H_{1,2}^{1,1}\left( k' \; \vline
\begin{array}{cc}
(0,1/2) & \\
(0,1/2), & (0,\gamma/2)
\end{array}
\right).
\end{equation}
The above Fourier cosine transforms can be solved by successive
applications of a Laplace and an inverse Laplace transform (a
technique pioneered by Fox \cite{fox2} for solving a wide variety
of integral transforms) to give \cite{srivastava}
\begin{eqnarray}
W(x,t) & = & \frac{1}{2 |x-a|} H_{3,3}^{2,1}\left(
\frac{|x-a|}{(Kt^{\gamma})^{1/2}} \; \vline
\begin{array}{ccc}
(1,1/2), & (1,\gamma/2), & (1,1/2)\\
(1,1), & (1,1/2), & (1,1/2)
\end{array}
\right) \nonumber \\
& &  - \frac{1}{2 (x+a)}
H_{3,3}^{2,1}\left( \frac{x+a}{(Kt^{\gamma})^{1/2}} \; \vline
\begin{array}{ccc}
(1,1/2), & (1,\gamma/2), & (1,1/2)\\
(1,1), & (1,1/2), & (1,1/2)
\end{array}
\right).
\end{eqnarray}

Now, applying Property 2 [cf. Eq. (\ref{p2})] of the H-functions
to reduce the order of our H-function, we get
\begin{eqnarray}
W(x,t) & = & \frac{1}{2 |x-a|} H_{2,2}^{2,0}\left(
\frac{|x-a|}{(Kt^{\gamma})^{1/2}} \; \vline
\begin{array}{cc}
(1,\gamma/2), & (1,1/2)\\
(1,1), & (1,1/2)
\end{array}
\right) \nonumber \\
& &  - \frac{1}{2 (x+a)}
H_{2,2}^{2,0}\left( \frac{x+a}{(Kt^{\gamma})^{1/2}} \; \vline
\begin{array}{cc}
(1,\gamma/2), & (1,1/2)\\
(1,1), & (1,1/2)
\end{array}
\right).
\end{eqnarray}
Applying Properties 1 and 3 [cf. Eq. (\ref{p3})] of
the H-functions, we can further reduce the order of our H-function:
\begin{eqnarray}
W(x,t) & = & \frac{1}{2 |x-a|} H_{1,1}^{1,0}\left(
\frac{|x-a|}{(Kt^{\gamma})^{1/2}} \; \vline
\begin{array}{c}
(1,\gamma/2)\\
(1,1)
\end{array}
\right) \nonumber \\
& &  - \frac{1}{2 (x+a)}
H_{1,1}^{1,0}\left( \frac{x+a}{(Kt^{\gamma})^{1/2}} \; \vline
\begin{array}{c}
(1,\gamma/2)\\
(1,1)
\end{array}
\right).
\end{eqnarray}
Applying Property 6 [cf. Eq. (\ref{p6})] of the H-functions
with $z=|x-a|$ (or $z=x+a$) and $\rho=-1$ we finally obtain
\begin{eqnarray}
&& W(x,t) = \nonumber \\
 && \frac{1}{2 (Kt^{\gamma})^{1/2}} \left[
H_{1,1}^{1,0}\left( \frac{|x-a|}{(Kt^{\gamma})^{1/2}} \; \vline
\begin{array}{c}
(1-\gamma/2,\gamma/2)\\
(0,1)
\end{array}
\right)- H_{1,1}^{1,0}\left( \frac{x+a}{(Kt^{\gamma})^{1/2}} \;
\vline
\begin{array}{c}
(1-\gamma/2,\gamma/2)\\
(0,1)
\end{array}
\right) \right]. \label{wkt}
\end{eqnarray}

Substituting Eq.~(\ref{wkt}) into Eq. (\ref{fT}) we have
\begin{eqnarray}
f(t) & = & -\frac{d}{dt} \left[ \frac{1}{2 (Kt^{\gamma})^{1/2}}
\int_0^{\infty} \ dx \
H_{1,1}^{1,0}\left( \frac{|x-a|}{(Kt^{\gamma})^{1/2}} \; \vline
\begin{array}{c}
(1-\gamma/2,\gamma/2)\\
(0,1)
\end{array}
\right) \right] \nonumber \\
& &  +\frac{d}{dT} \left[ \frac{1}{2 (Kt^{\gamma})^{1/2}}
\int_0^{\infty} \
dx \
H_{1,1}^{1,0}\left( \frac{x+a}{(Kt^{\gamma})^{1/2}} \; \vline
\begin{array}{c}
(1-\gamma/2,\gamma/2)\\
(0,1)
\end{array}
\right)
\right].
\end{eqnarray}
Defining $z=(x-a)/(KT^{\gamma})^{1/2}$,
$z'=(x+a)/(KT^{\gamma})^{1/2}$, we obtain
\begin{eqnarray}
f(t) & = & -\frac{d}{dt} \int_{-a/(Kt^{\gamma})^{1/2}}^{\infty} \ dz \
H_{1,1}^{1,0}\left( |z| \; \vline
\begin{array}{c}
(1-\gamma/2,\gamma/2)\\
(0,1)
\end{array}
\right) \nonumber \\
& & + \frac{d}{dt} \int_{a/(Kt^{\gamma})^{1/2}}^{\infty} \ dz' \
H_{1,1}^{1,0}\left( z' \; \vline
\begin{array}{c}
(1-\gamma/2,\gamma/2)\\
(0,1)
\end{array}
\right).
\end{eqnarray}
Evaluating the above equation (which is easily done since the only
$t$ dependence is in the limits of the integrals), we obtain the following
expression for the first passage time density
\begin{equation}
f(t) = \frac{a \gamma}{2 K^{1/2} t^{(2+\gamma)/2}}
H_{1,1}^{1,0}\left( \frac{a}{(Kt^{\gamma})^{1/2}} \; \vline
\begin{array}{c}
(1-\gamma/2,\gamma/2)\\
(0,1)
\end{array}
\right). \label{sol4}\end{equation} We mention that this result
has appeared earlier in a short communication \cite{rd}. It should
be noted that H-functions were first used in the context of
probability distributions by Schneider \cite{schneider}. They have
also been used to express solutions of fractional diffusion
equations \cite{schneider2}. The series expansion of the
H-function in Eq.~(\ref{sol4}) [cf. Eq. (\ref{hfn})] is
\begin{equation}
f(t) = \frac{a \gamma}{2 K^{1/2} t^{(2+\gamma)/2}}
\sum_{k=0}^{\infty} \frac{(-a/(Kt^{\gamma})^{1/2})^k}{k!
\Gamma(1-\gamma/2-k \gamma/2)}.
\label{ftexp}\end{equation}
This turns out to be also the series expansion of the Maitland's
generalized hypergeometric function or the Wright function $_0
\psi_1$ \cite{erdelyi}. Thus, an alternative expression for $f(t)$
is
\begin{equation}
f(t) = \frac{a \gamma}{2 K^{1/2} t^{(2+\gamma)/2}} \
_0 \psi_1 \left(
\begin{array}{c}
- \\
(1-\gamma/2,-\gamma/2)
\end{array}; -\frac{a}{(Kt^{\gamma})^{1/2}}
\right). \label{main}
\end{equation}

It is difficult to work with the series expansion of $f(t)$ given
in Eq. (\ref{ftexp}) due to the presence of the gamma function with
large negative arguments. We get around this as follows.
From the properties of gamma function \cite{gradshteyn} we have
\be
\Gamma(1-\gamma/2-k \gamma/2) = - \frac{(k+1)\gamma}{2} \Gamma(-(k+1)\gamma /2).
\label{gamneg}\ee
Using the following relation \cite{gradshteyn}
\be
\Gamma(-z)= \frac{-\pi}{\sin(\pi z)\Gamma(z+1)},
\ee
Eq. (\ref{gamneg}) can be rewritten as
\bea
\Gamma(1-\gamma/2-k \gamma/2) & = & \frac{\pi (k+1)\gamma}{2}
\frac{1}{\sin[(k+1)\pi \gamma/2] \Gamma[1+(k+1)\gamma /2]} \nonumber \\
 & = & \frac{\pi}{\sin[(k+1)\pi \gamma/2] \Gamma[(k+1)\gamma /2]}.
\eea
Substituting this in Eq. (\ref{ftexp}), we obtain
\begin{equation}
f(t) = \frac{a \gamma}{2 \pi K^{1/2} t^{(2+\gamma)/2}}
\sum_{k=0}^{\infty}
\frac{(-a/(Kt^{\gamma})^{1/2})^k \sin[(k+1)\pi \gamma/2] \Gamma[(k+1)\gamma /2]}{\Gamma(k+1)}.
\label{ftexp2}\end{equation}

Note that for regular Brownian motion ($\gamma=1$), the above
expression for $f(t)$ reduces to
\begin{equation}
f(t) = \frac{a}{2 \pi \sqrt{Kt^3}}
\sum_{k=0}^{\infty}
\frac{(-a/\sqrt{Kt})^k \sin[(k+1)\pi /2] \Gamma[(k+1)/2]}{\Gamma(k+1)}.
\label{rbm}\end{equation}
But \cite{gradshteyn}
\be
\Gamma(k+1) = \frac{4^{k/2} \Gamma[(k+1)/2] \Gamma(1+k/2)}{\sqrt{\pi}}.
\ee
Further,
\begin{eqnarray}
\sin[(k+1) \pi /2] & = & 0, \ \ k \ {\rm odd}, \nonumber \\
& = & (-1)^{k/2} \ \ k\  {\rm even}.
\end{eqnarray}
Substituting these relations in Eq. (\ref{rbm}) and letting $n=k/2$ we get
\begin{equation}
f(t) = \frac{a}{2 \pi \sqrt{Kt^3}} \sum_{n=0}^{\infty}
\frac{(-a/\sqrt{Kt})^{2n} \Gamma[(2n+1)/2] (-1)^n}{(4)^{n} \Gamma[(2n+1)/2 \Gamma(n+1)}.
\end{equation}
Simplifying this we finally obtain
\begin{eqnarray}
f(t) & = & \frac{a}{\sqrt{4 \pi Kt^3}} \sum_{n=0}^{\infty}
\frac{(-a^2/4\sqrt{Kt})^{n}}{n!} \nonumber \\
& = & \frac{a}{\sqrt{4 \pi Kt^3}} \exp[-a^2/4 Kt].
\end{eqnarray}
This is the expected inverse Gaussian distribution
for the FPT density function of the ordinary Brownian motion with
zero drift \cite{grimmet}.

Next we consider the asymptotic behavior of the FPT
distribution for large values of $t$. Refer to Eq.~(\ref{sol4}).
Let $z=a/(Kt^{\gamma})^{1/2}$. It is known that
\cite{mathai,braaksma}, for small $z$, $H_{1,1}^{1,0}(z) \sim
|z|^{b_1/B_1}=1$, since $b_1=0$ and $B_1=1$. Therefore, the FPT
distribution $f(t)$, for large $t$, is characterized by the power
law relation
\begin{equation}
f(t) \sim t^{-1-\gamma/2}, \label{universal}
\end{equation}
which becomes the well known $-3/2$ scaling law for the ordinary
Brownian motion. We make two comments.
First, the above power law behavior has been observed earlier
by Balakrishnan \cite{balakrishnan} for subdiffusive processes ($0
< \gamma < 1$) using a different method. Using our method the same
scaling law is shown to be applicable also to superdiffusive
processes. Second,
from Eq.~(\ref{universal}), we see that the mean first passage
time and all higher moments of the FPT distribution are undefined
for $0 < \gamma < 2$.

Next we consider the asymptotic behavior of $f(t)$ for small $t$
(i.e. large $z$ where $z=a/(Kt^{\gamma})^{1/2}$). It is known
\cite{mathai,braaksma} that for large $z$,
\be
H^{1,0}_{1,1}(z) \sim B z^{\alpha/\mu} \exp[- \mu C^{1/\mu} z^{1/\mu}],
\ee
where
\ben
\alpha & = & \sum_{j=1}^q b_j - \sum_{j=1}^p a_j + (p-q+1)/2, \\
C & = & \prod_{j=1}^p (A_j)^{A_j} \prod_{l=1}^q (B_j)^{B_j}, \\
\mu & = & \sum_{j=1}^q B_j - \sum_{j=1}^p A_j, \\
B & = & (2 \pi)^{(q-p-1)/2} C^{\alpha/\mu} \mu^{-1/2}
\prod_{j=1}^p (A_j)^{-a_j+0.5} \prod_{l=1}^q (B_j)^{b_j-0.5}.
\een
In our case, $p=1$, $q=1$, $a_1=1-\gamma/2$, $A_1=\gamma/2$,
$b_1=0$ and $B_1=1$. Therefore, for $z$ large
\bea
H_{1,1}^{1,0}\left( \frac{a}{(Kt^{\gamma})^{1/2}} \; \vline
\begin{array}{c}
(1-\gamma/2,\gamma/2)\\
(0,1)
\end{array}
\right) & \sim & \frac{1}{\sqrt{(2-\gamma) \pi}}
\left( \frac{\gamma}{2} \right)^{(\gamma-1)/(2-\gamma)}
z^{(\gamma-1)/(2-\gamma)} \nonumber \\
 & & \exp \left[ -\frac{2-\gamma}{2}
\left( \frac{\gamma}{2} \right)^{\gamma/(2-\gamma)}
z^{2/(2-\gamma)} \right].
\eea
Substituting this in the expression for $f(t)$ [cf. Eq. (\ref{sol4})],
we obtain for small $t$
\be
f(t) \sim \frac{r}{t^{(4-\gamma)/(4-2\gamma)}} \exp \left[ -
\frac{d}{t^{\gamma/(2-\gamma)}} \right], \label{ftsmall}\ee where
\ben r & = & \frac{a \gamma}{\sqrt{4 K \pi (2-\gamma)}} \left(
\frac{a \gamma}{2 \sqrt{K}} \right)^{(\gamma-1)/(2-\gamma)}, \\ d
& = & \frac{2-\gamma}{2} \left( \frac{\gamma}{2}
\right)^{\gamma/(2-\gamma)} \left( \frac{a}{\sqrt{K}}
\right)^{2/(2-\gamma)}. \een For $0 < \gamma < 2$, both $r$ and
$d$ are greater than zero. Therefore, $f(t)$ is exponentially
decaying for small $t$. Note that for $\gamma=1$ (ordinary
Brownian motion), the above expression reduces to
\be
f(t) \sim \frac{a}{\sqrt{4 \pi Kt^3}} \exp[-a^2/4 Kt].
\ee
In this case, the asymptotic expression turns out to be the
exact expression.

We can determine the $t$ value where $f(t)$ attains its maximum
value from the above expression. We obtain
\be
t_{\rm max} = \left( \frac{2 d \gamma}{4-\gamma}
\right)^{(2-\gamma)/\gamma}. \ee For ordinary Brownian motion
($\gamma =1$),
\be
t_{\rm max} = \frac{2d}{3} = \frac{a^2}{6K}.
\ee

The theoretical prediction for the full FPT density function
given in Eq. (\ref{sol4}) is verified by numerically simulating the underlying CTRW process
characterized by the probability density function $\phi(y,u)$ [cf.
Eq. (\ref{LWpdf})]. For the sake of numerical efficiency, we
replace the waiting time distribution $(\alpha-1)/\tau
(1+u/\tau)^{\alpha}$ in $\phi(y,u)$ by the Pareto density function
\cite{evans} (which is equal to zero for $u < \tau$ and
$(\alpha-1) \tau^{\alpha-1}/u^{\alpha}$ for $u \ge \tau$).
The parameter values are
chosen as follows: $\gamma=0.5$, $a=1.0$ and $K=0.1$.
In Figure 2,
the FPT density function obtained theoretically from Eq.
(\ref{sol4}) is compared with the FPT density function obtained
numerically using 10 million realizations of the underlying
stochastic process. If we take
a large value for $\tau$ (=0.01), the agreement is not that good (see
Figure 2a) since we are not yet close to the generalized diffusion
limit. If, however, we take $\tau$ to be $10^{-4}$, the numerical simulation is in
excellent agreement with the theoretical prediction (see Figure 2b).  The agreement
gets better as $\tau$ and $\sigma$ become smaller, approaching the
generalized diffusion limit.

The above statement can be quantified as follows.
Consider the Kullback-Leibler information criterion \cite{kullback,white} which
gives a quantitative measure of how ``far apart'' a given
approximate density function $f_1(t)$ is from the exact density function
$f(t)$. The Kullback-Leibler information criterion is defined as
\be
KL(f,f_1) = \int_0^{\infty} dt f(t) \log \frac{f(t)}{f_1(t)}.
\ee
This is always greater than or equal to zero and is equal to zero
if and only if $f_1(t)$ agrees exactly with $f(t)$.
The deviation away from zero quantifies the disagreement between
the two density functions.

In our case, we take $f_1(t)$ to be
the approximate numerically simulated FPT density function
for various values of $\tau$ (the value of $\sigma$ is correspondingly
varied to keep $K$ constant). The plot of the Kullback-Leibler
information criterion for different values of $1/\tau$ is given
in Figure 3. As $\tau$ decreases (i.e., as the generalized diffusion
limit is approached), the Kullback-Leibler information criterion
also decreases.

\subsection{Nonzero Drift Case}

In this section, we consider the first passage time problem for
Levy type anomalous diffusion with drift. An explicit solution for
the FPT density function
is not possible in this case. We obtain the Laplace transform
of the FPT density function (the so-called ``moment generating
function'' \cite{cox}). Using this, we obtain the mean and variance
of the first passage time distribution. For regular Brownian motion
($\gamma=1$), the inverse Laplace transform of the moment generating
function can be explicitly carried out to give the standard results.

Taking the Laplace transform of Eq. (\ref{FFPE2}) we get
\begin{equation}
q(x,s)-\frac{W(x,0)}{s} = \frac{K}{s^{\gamma}} \frac{\partial^2}{\partial x^2}
q(x,s) - \frac{\mu}{s} \frac{\partial}{\partial x} q(x,s),
\end{equation}
where $q(x,s)$ is the Laplace transform of $W(x,t)$.
Here we have again applied the result \cite{miller} that the Laplace
transform of $_0 D_t^{-\gamma} W(x,t)$ is $q(x,s)/s^{\gamma}$. The
above equation can be rewritten as
\begin{equation}
\frac{\partial^2}{\partial x^2}q(x,s) + A \frac{\partial}{\partial x} q(x,s)
+ B q(x,s) = -\frac{s^{\gamma-1}}{K} \delta(x-a),
\label{diffeq}\end{equation}
where
\begin{equation}
A = -\frac{\mu s^{\gamma-1}}{K}; \ \ \ B = -\frac{s^{\gamma}}{K}.
\label{const}\end{equation}

Since $s,K >0$, we have
\begin{equation}
\lambda^2 \equiv A^2-4B = \frac{\mu^2 s^{2\gamma-2}}{K^2} +
4 \frac{s^{\gamma}}{K} > 0.
\end{equation}
Therefore two independent solutions of the homogeneous equation
corresponding to
Eq. (\ref{diffeq}) are given by \cite{polyanin}
\begin{equation}
q_1(x,s) = \exp[x(\lambda-A)/2]; \ \ \ q_2(x,s) = \exp[x(-\lambda-A)/2].
\end{equation}
Consequently, the general solution of Eq. (\ref{diffeq}) satisfying all the
boundary and initial conditions [cf. Eq. (\ref{bc2})] is given by
\begin{equation}
q(x,s) = \frac{s^{\gamma-1}}{K \lambda}e^{-A(x-a)/2}[e^{-\lambda|x-a|/2} -
e^{-\lambda(x+a)/2}].
\label{ltp}\end{equation}

To obtain the Laplace transform of the FPT density function, we
take the Laplace transform of Eq. (\ref{fT}) to get
\begin{equation}
F(s) = -s \int_0^{\infty} dx \, q(x,s) + \int_0^{\infty} dx \, W(x,0).
\end{equation}
Here we have used the fact that Laplace transform of $dW(x,t)/dt$ is
given by \cite{erdelyi2} $s q(x,s) - W(x,0)$. Since $W(x,0) = \delta(x-a)$
[cf. Eq. (\ref{bc2})], we obtain
\begin{equation} \label{fslt}
F(s) = 1-s \int_0^{\infty} dx \, q(x,s).
\end{equation}
Substituting for $q(x,s)$ from Eq. (\ref{ltp}), we get
\begin{eqnarray*}
F(s) &=& 1-\frac{s^{\gamma}}{K \lambda} \left[\int_0^a dx \,
e^{-A(x-a)/2} e^{-\lambda(a-x)/2} - \int_a^{\infty} dx \,
e^{-A(x-a)/2} e^{-\lambda(x-a)/2} \right] \\
&+& \frac{s^{\gamma}}{K \lambda} \int_0^{\infty}
dx \, e^{-A(x-a)/2} e^{-\lambda(x+a)/2}.
\end{eqnarray*}
The integrals can be easily evaluated to finally give [upon using
Eq. (\ref{const})]
\begin{equation}\label{ltfpt}
F(s) = \exp\left[-\frac{a \mu s^{\gamma-1}}{2K}-
a \sqrt{\frac{\mu^2 s^{2\gamma-2}}{4K^2} + \frac{s^{\gamma}}{K}} \right].
\end{equation}

For $0 < \gamma < 1$, $F(s)$ is not a completely monotone
function \cite{feller}. That is, it does not satisfy the following
conditions
\be \label{monotone} (-1)^n \frac{d F^n(s)}{ds^n} \ge 0
\ {\rm for\ } s \ge 0; \ \ F(0) = 1.
\ee
Hence $F(s)$ is not a
Laplace transform of a probability density function. The
physical reason for this is not yet understood. For $1 \le
\gamma < 2$, we have not been able to prove rigorously that
$F_{\gamma,\mu}(s)$ is a completely monotone function. However,
we have calculated the first hundred derivatives of
$F_{\gamma,\mu}(s)$ using the symbolic manipulation program
Mathematica and find that all of them satisfy Eq.
(\ref{monotone}). Hence we conjecture that $F(s)$ is the Laplace
transform of a probability density function for $1 \le \gamma <
2$. Henceforth, we restrict ourselves to this parameter range.

First consider the case where the drift is
towards the barrier. This implies that $\mu < 0$ since in our
formulation the diffusive process starts at $x=a>0$ and the barrier is at $x=0$.
In this case, the FPT density function can be written as
[cf. Eq. (\ref{ltfpt})]
\begin{equation}
F(s) = e^{a g(s)},
\label{ltfpt2}\end{equation}
where
\begin{equation}
g(s) = \frac{|\mu| s^{\gamma-1}}{2K} - \frac{|\mu| s^{\gamma-1}}{2K}
\sqrt{1+\frac{4K s^{2-\gamma}}{\mu^2}}.
\label{gs}\end{equation}

The mean first passage time is given by
\begin{equation}
\langle T \rangle = - \frac{d F(s)}{d s} \vline_{s=0}.
\label{mean}\end{equation}
From Eqs. (\ref{ltfpt2}) and (\ref{gs}), we have
\begin{eqnarray}\label{mom1}
&&\frac{d F(s)}{d s} = \nonumber \\ && \left[
\frac{|\mu|(\gamma-1)s^{\gamma-2}a}{2K} \left(1- \sqrt{1+\frac{4K
s^{2-\gamma}}{\mu^2}}\right) - \frac{(2-\gamma)a(1+\frac{4K
s^{2-\gamma}}{\mu^2})^{-1/2}}{|\mu|} \right] e^{a g(s)}.
\end{eqnarray}
We need to find the limiting value of
the above expression as $s \rightarrow 0$. First consider $e^{a
g(s)}$. As $s \rightarrow 0$, we can expand the square root in Eq.
(\ref{gs}) to give
\begin{equation}
g(s) = - \frac{s}{|\mu|} + \frac{K s^{3-\gamma}}{|\mu|^3} - \cdots
\end{equation}
Hence $g(s) \rightarrow 0$ as $s \rightarrow 0$. Consequently,
$e^{a g(s)} \rightarrow 1$ as $s \rightarrow 0$. Performing similar
expansions for the other terms in Eq. (\ref{mom1}), we finally obtain
[cf. Eq. (\ref{mean})]
\begin{equation}
\langle T \rangle = \frac{a}{|\mu|}.
\end{equation}
Note that the mean first passage time is independent of $\gamma$.

The variance is obtained as follows:
\begin{equation}\label{var}
{\rm Var}(T) = \langle T^2 \rangle - \langle T \rangle^2.
\end{equation}
Therefore we need to evaluate $\langle T^2 \rangle$. This is given by
\begin{equation}
\langle T^2 \rangle = \frac{d^2 F(s)}{d s^2} \vline_{s=0}.
\label{mom2}\end{equation}
Now
\begin{equation}
\frac{d^2 F(s)}{d s^2} = \left[ a \frac{d^2 g(s)}{ds^2} +
\left( a \frac{d g(s)}{ds} \right)^2 \right] e^{a g(s)}.
\label{secder}\end{equation}

Consider the first term. The second derivative of $g(s)$ is given by
\begin{eqnarray*}
\frac{d^2 g(s)}{ds^2} & = & \frac{|\mu|(\gamma-1)(\gamma-2)s^{\gamma-3}}{2K}\left(1-
\sqrt{1+\frac{4K s^{2-\gamma}}{\mu^2}}\right) \nonumber \\
& & -\frac{(\gamma-1)(2-\gamma)}{|\mu|s}
\left(1+\frac{4K s^{2-\gamma}}{\mu^2}\right)^{-1/2} \\
& & + \frac{2 K (2-\gamma)^2 s^{1-\gamma}}{|\mu|^3}
\left(1+\frac{4K s^{2-\gamma}}{\mu^2}\right)^{-3/2}. \nonumber
\end{eqnarray*}
Expanding all terms we get
\begin{eqnarray*}
\frac{d^2 g(s)}{ds^2} & = & -\frac{|\mu|(\gamma-1)(\gamma-2)s^{\gamma-3}}{2K}
\sum_{n=2}^{\infty} \frac{(-1)^{n-1} 1 \cdot 3 \cdots (2n-3)}{2^n n!}
\left(\frac{4K s^{2-\gamma}}{\mu^2}\right)^n \nonumber \\
& & -\frac{(\gamma-1)(2-\gamma)}{|\mu|s}
\sum_{n=1}^{\infty} \frac{(-1)^{n} 1 \cdot 3 \cdots (2n-1)}{2^n n!}
\left(\frac{4K s^{2-\gamma}}{\mu^2}\right)^n \\
& & + \frac{2 K (2-\gamma)^2 s^{1-\gamma}}{|\mu|^3}
\sum_{n=0}^{\infty} \frac{(-1)^{n} 1 \cdot 3 \cdots (2n+1)}{2^n n!}
\left(\frac{4K s^{2-\gamma}}{\mu^2}\right)^n. \nonumber
\end{eqnarray*}
After considerable manipulation, this can be rewritten as
\begin{eqnarray*}
\frac{d^2 g(s)}{ds^2} & = &  \frac{2 K (2-\gamma) s^{1-\gamma}}{|\mu|^3}
\sum_{n=0}^{\infty} \left[ \frac{n(2-\gamma)+(3-\gamma)}{(n+2)} \right] \\
& & \frac{(-1)^{n} 1 \cdot 3 \cdots (2n+1)}{2^n n!}
\left(\frac{4K s^{2-\gamma}}{\mu^2}\right)^n.
\end{eqnarray*}
Thus the first term in Eq. (\ref{secder}) is given by
\begin{eqnarray}\label{1stterm}
a \frac{d^2 g(s)}{ds^2} e^{ag(s)} & = &  \frac{2 K a (2-\gamma) s^{1-\gamma}}{|\mu|^3}
e^{ag(s)} \sum_{n=0}^{\infty} \left[ \frac{n(2-\gamma)+(3-\gamma)}{(n+2)} \right]  \\
& & \frac{(-1)^{n} 1 \cdot 3 \cdots (2n+1)}{2^n n!}
\left(\frac{4K s^{2-\gamma}}{\mu^2}\right)^n. \nonumber
\end{eqnarray}
For $\gamma=1$, the case of ordinary Brownian motion, we obtain from
the above equation
\begin{equation}
a \frac{d^2 g(s)}{ds^2} e^{ag(s)} \vline_{s=0} = \frac{\sigma^2 a}{|\mu|^3}.
\end{equation}
Here we have also used the fact that $K=\sigma^2/2$ for $\gamma=1$.
On the other hand, for $1 < \gamma < 2$, as $s \rightarrow 0$ the
prefactor multiplying the sum in Eq. (\ref{1stterm}) diverges whereas
the sum itself is finite and bounded away from zero. Consequently, for
$1 < \gamma < 2$ the first term in Eq. (\ref{secder}) diverges as
$s \rightarrow 0$.

Next, we consider the second term in Eq. (\ref{secder}). We
already know the limiting behaviour of this term for
$1 \le \gamma < 2$ as $s \rightarrow 0$
from our earlier analysis for mean first passage
time, namely,
\begin{equation}
\left(\frac{d g(s)}{ds} \right)^2 e^{ag(s)} \vline_{s=0} = \frac{a^2}{\mu^2}.
\end{equation}
Substituting the above results in Eq. (\ref{mom2}), for $\gamma=1$ we obtain
\begin{equation}
\langle T^2 \rangle = \frac{\sigma^2 a}{|\mu|^3} + \frac{a^2}{\mu^2}.
\end{equation}
Therefore [cf. Eq. (\ref{var})]
\begin{equation}
{\rm Var}(T) = \sigma^2 a/|\mu|^3,
\end{equation}
for $\gamma=1$. For $1 < \gamma < 2$, the first term in Eq. (\ref{secder}) diverges
as $s \rightarrow 0$ whereas the second term is finite. Therefore
$\langle T^2 \rangle$ diverges. Hence the variance also diverges.

Next, consider the case where the drift is away from the barrier.
This implies that $\mu > 0$ in our formulation. Now the FPT
density function can be written as in Eq. (\ref{ltfpt2}) where
\begin{equation}
g(s) = - \frac{|\mu| s^{\gamma-1}}{2K} - \frac{|\mu| s^{\gamma-1}}{2K}
\sqrt{1+\frac{4K s^{2-\gamma}}{\mu^2}}.
\end{equation}
Performing the same analysis as above, it is easily seen that the
mean and variance diverge for $1 \le \gamma < 2$.

For $\gamma=1$, the Laplace transform of the
first passage time density function reduces to [cf. Eq.
(\ref{ltfpt})]
\begin{equation}
F(s) = \exp\left[-\frac{a}{2 K}( \mu + \sqrt{\mu^2
+ 4 K s} \right].
\label{ltfptB}\end{equation}
Now the
inverse Laplace transform of $\exp(-\alpha \sqrt{s}), \ \alpha \ge
0$ is \cite{erdelyi2}
\begin{equation}
\frac{\alpha}{2 \sqrt{\pi t^3}} \exp(-\alpha^2/4t).
\end{equation}
Using this result, we can easily perform the inverse Laplace transform of
Eq. (\ref{ltfptB}) to obtain
\begin{equation}
f(t) = \frac{a}{\sqrt{4 \pi K t^3}} \exp \left[ -
\frac{(a+\mu t)^2}{4 K t} \right], \ \ \ a >0, \ \ t > 0.
\end{equation}
We comment that, if the starting point of the diffusion is chosen
at $x(0)=0$, a negative $\mu$ will be used in the above equation
and we recover the expected inverse Gaussian density \cite{grimmet}
for the FPT density function of Brownian
motion with drift .

\subsection{Absorbing Barriers at $x=-b$ and $x=a$}

In this section, we consider the first passage time problem
for Levy type anomalous diffusion with zero drift and absorbing barriers placed at $x=-b$
and $x=a$.  We obtain an explicit solution for the
first passage time density.

The FFPE to be solved is given in Eq. (\ref{FFPE}). The boundary and initial
conditions become:
\be
W(-b,t) = W(a,t) = 0, \ \ \ W(x,0) = \delta(x).
\label{bc3}\ee
We solve the FFPE using the method of separation of variables \cite{morse}.
Let $W(x,t)=X(x) T(t)$. Substituting in Eq. (\ref{FFPE}) we obtain
\be
X(x)T(t) - X(x) = \ _0D_t^{-\gamma} T(t) K X''(x),
\ee
where $X''(x)$ denotes the second derivative of $X(x)$ with respect to $x$.
Separating out the variables and introducing the separation constant $\lambda$
we get
\be
K X''(x) =  \lambda X(x),
\label{xeq}\ee
and
\be
T(t)-1  =  \lambda \ _0D_t^{-\gamma} T(t).
\label{teq}\ee

The solution of Eq. (\ref{xeq}) with the given boundary conditions
is given by
\be
X_n(x) = A_n \sin \left[ \frac{n \pi (b+x)}{(a+b)} \right].
\ee
with
\be
\lambda_n = - \frac{n^2 \pi^2}{(a+b)^2} K, \ \ \ n=1,2, \ldots\ .
\ee
To solve Eq. (\ref{teq}), we take its Laplace transform to obtain
(introducing the subscript $n$ coming from $\lambda_n$)
\be
T_n(s)-\frac{1}{s} = \frac{\lambda_n}{s^{\gamma}} T_n(s).
\ee
Here we have used the fact that the Laplace transform of
$ _0D_t^{-\gamma} T(t)$ is $T(s)/s^{\gamma}$. Solving for $T_n(s)$ we get
\be
T_n(s) = \frac{1}{s-\lambda_n s^{1-\gamma}}.
\ee
Taking the inverse Laplace transform \cite{erdelyi} we finally obtain
\be
T_n(t) = E_{\gamma}\left[ -\frac{n^2 \pi^2}{(a+b)^2} K t^{\gamma} \right],
\ee
where $E_{\gamma}(z)$ is the Mittag-Leffler function introduced earlier
[cf. Eq. (\ref{mf})].

Combining the solutions for space and time parts, we get
\be
W(x,t) = \sum_{n=1}^{\infty} A_n \sin \left[ \frac{n \pi (b+x)}{(a+b)} \right]
E_{\gamma}\left[ -\frac{n^2 \pi^2}{(a+b)^2} K t^{\gamma} \right].
\ee
The coefficients $A_n$ are determined by imposing the initial condition
$W(x,0)=\delta(x)$. This gives us
\be
A_n = \frac{2}{a+b} \sin \left[ \frac{n \pi b}{(a+b)} \right].
\ee
Hence
\be
W(x,t) = \sum_{n=1}^{\infty} \frac{2}{a+b} \sin \left[ \frac{n \pi b}{(a+b)} \right]
 \sin \left[ \frac{n \pi (b+x)}{(a+b)} \right]
E_{\gamma}\left[ -\frac{n^2 \pi^2}{(a+b)^2} K t^{\gamma} \right].
\ee

From Eq. (\ref{fT}), the FPT density function is given by
\begin{equation}
f(t) = - \frac{d}{dt} \int_{-b}^{a}\ dx \ W(x,t).
\end{equation}
Substituting for $W(x,t)$ into this equation, we get
\be
f(t) = -\frac{d}{dt} \left\{4 \sum_{n=0}^{\infty} \frac{1}{(2n+1)\pi}
\sin \left[ \frac{(2n+1) \pi b}{(a+b)} \right]
E_{\gamma}\left[ -\frac{(2n+1)^2 \pi^2}{(a+b)^2} K t^{\gamma} \right] \right\}.
\label{ftder}\ee
Here we have used the fact that
\ben
\int_{-b}^a \sin \left[ \frac{n \pi (b+x)}{(a+b)} \right] & = &
\frac{2 (a+b)}{n \pi} \ \ {\rm if\ } n \ {\rm is\ odd}, \\
 & = & 0  \ \ {\rm if\ } n \ {\rm is\  even}.
\een

To evaluate the derivative in Eq. (\ref{ftder}), we write the
Mittag-Leffler function in terms of the H-function using
Eq. (\ref{mfhfn}):
\be
f(t) = - \frac{4}{\pi} \sum_{n=0}^{\infty} \frac{1}{2n+1}
\sin \left[ \frac{(2n+1) \pi b}{(a+b)} \right]
\frac{d}{dt} H_{1,2}^{1,1}\left(\frac{(2n+1)^2 \pi^2}{(a+b)^2} K t^{\gamma} \; \vline
\begin{array}{cc}
(0,1) & \\
(0,1), & (0,\gamma)
\end{array}
\right).
\ee
But \cite{mathai}
\ben
\frac{d}{dt} H_{1,2}^{1,1}\left(\frac{(2n+1)^2 \pi^2}{(a+b)^2} K t^{\gamma} \; \vline
\begin{array}{cc}
(0,1) & \\
(0,1), & (0,\gamma)
\end{array}
\right) & = &
\frac{1}{t} H_{2,3}^{1,2}\left(\frac{(2n+1)^2 \pi^2}{(a+b)^2} K t^{\gamma} \; \vline
\begin{array}{ccc}
(0,\gamma) & (0,1) &  \\
(0,1), & (0,\gamma) & (1,\gamma)
\end{array}
\right) \\
& = & \frac{1}{t} H_{1,2}^{1,1}\left(\frac{(2n+1)^2 \pi^2}{(a+b)^2} K t^{\gamma} \; \vline
\begin{array}{cc}
(0,1) & \\
(0,1), & (1,\gamma)
\end{array}
\right).
\een
In the last step, we have used properties 1 and 2 of H-functions [cf. Appendix].
From the series expansion of the H-function given in Eq. (\ref{hfn}), it can
be shown after some manipulation that
\be
H_{1,2}^{1,1}\left(z \; \vline
\begin{array}{cc}
(0,1) & \\
(0,1), & (1,\gamma)
\end{array}
\right) =
-z E_{\gamma,\gamma}(-z),
\ee
where $E_{\alpha,\beta}(z)$ is the generalized Mittag-Leffler function \cite{erdelyi}
\be
E_{\alpha,\beta}(z) = \sum_{k=0}^{\infty} \frac{z^k}{\Gamma(\beta+k \alpha)}, \ \
\alpha, \beta > 0.
\ee
Using this in the expression for $f(t)$ we get
\be
f(t) = \frac{4 \pi K t^{\gamma -1}}{(a+b)^2}
\sum_{n=0}^{\infty} (2n+1)
\sin \left[ \frac{(2n+1) \pi b}{(a+b)} \right]
E_{\gamma,\gamma}\left[ -\frac{(2n+1)^2 \pi^2}{(a+b)^2} K t^{\gamma} \right].
\ee
For a regular Brownian motion ($\gamma=1$), we obtain
\be
f(t) = \frac{4 \pi K}{(a+b)^2}
\sum_{n=0}^{\infty} (2n+1)
\sin \left[ \frac{(2n+1) \pi b}{(a+b)} \right]
\exp\left[ -\frac{(2n+1)^2 \pi^2}{(a+b)^2} K t  \right].
\ee

\section{Conclusions}

We studied the first passage time (FPT) problem for Levy type anomalous
diffusion. We obtained the FPT distribution using the recently formulated
fractional Fokker-Planck equation. We derived an explicit expression for
the FPT distribution in terms of Fox or H-functions when the diffusion
has zero drift. The theoretical result was verified by numerically
simulating the underlying continuous time random walk. When the drift is
nonzero, we obtained an analytic expression for the Laplace transform of
the FPT distribution. This was used to calculate the mean and variance of
the FPT distribution. Finally, for the case of two absorbing
barriers at finite distances from the origin we expressed the FPT distribution in terms of
a power series. In all of the above situations, the known results for ordinary
diffusion (Brownian motion) were obtained as special cases of our
more general results.

\newpage
\section*{Acknowledgements}

This work was supported by US ONR Grant N00014-99-1-0062. GR
thanks Center for Complex Systems and Brain Sciences, Florida
Atlantic University, where this work was performed, for
hospitality.

\newpage
\appendix
\section*{Properties of H-functions}

The H-function has the following remarkable properties
\cite{mathai} which we will use later.

{\em Property 1.} The H-function is symmetric in the pairs
$(a_1,A_1), \ldots ,(a_n,A_n)$, likewise $(a_{n+1},A_{n+1}),
\ldots ,(a_p,A_p)$; in $(b_1,B_1), \ldots ,(b_m,B_m)$ and in
$(b_{m+1},B_{m+1}), \ldots ,(b_q,B_q)$.

{\em Property 2.} Provided $n \geq 1$ and $q > m$,
\begin{eqnarray} \label{p2}
&& H_{p,q}^{m,n}\left( z \; \vline
\begin{array}{cccc}
(a_1,A_1), & (a_2,A_2), & \cdots , & (a_p,A_p)\\
(b_1,B_1), & \cdots , & (b_{q-1},B_{q-1}), & (a_1,A_1)
\end{array}
\right) \nonumber \\ & = & H_{p-1,q-1}^{m,n-1}\left( z \; \vline
\begin{array}{cccc}
(a_2,A_2), & \cdots , & (a_p,A_p)\\
(b_1,B_1), & \cdots , & (b_{q-1},B_{q-1})
\end{array}
\right).
\end{eqnarray}

{\em Property 3.} Provided $m \geq 2$ and $p > n$,
\begin{eqnarray}\label{p3}
&& H_{p,q}^{m,n}\left( z \; \vline
\begin{array}{cccc}
(a_1,A_1), & \cdots , & (a_{p-1},A_{p-1}), & (b_1,B_1)\\
(b_1,B_1), & (b_2,B_2), & \cdots , & (b_q,B_q)
\end{array}
\right) \nonumber \\
 & = & H_{p-1,q-1}^{m-1,n}\left( z \; \vline
\begin{array}{cccc}
(a_1,A_1), & \cdots , & (a_{p-1},A_{p-1})\\
(b_2,B_2), & \cdots , & (b_q,B_q)
\end{array}
\right).
\end{eqnarray}

{\em Property 4.}
\begin{equation}\label{p4}
H_{p,q}^{m,n}\left( z \; \vline
\begin{array}{c}
(a_j,A_j)_{j=1, \ldots ,p}\\
(b_j,B_j)_{j=1, \ldots ,q}
\end{array}
\right) = H_{q,p}^{n,m}\left( \frac{1}{z} \; \vline
\begin{array}{c}
(1-b_j,B_j)_{j=1, \ldots ,q} \\
(1-a_j,A_j)_{j=1, \ldots ,p}
\end{array}
\right).
\end{equation}

{\em Property 5.} For $k > 0$,
\begin{equation}\label{p5}
\frac{1}{k} H_{p,q}^{m,n}\left( z \; \vline
\begin{array}{c}
(a_j,A_j)_{j=1, \ldots ,p}\\
(b_j,B_j)_{j=1, \ldots ,q}
\end{array}
\right) = H_{p,q}^{m,n}\left( z^k \; \vline
\begin{array}{c}
(a_j,k A_j)_{j=1, \ldots ,p}\\
(b_j,k B_j)_{j=1, \ldots ,q}
\end{array}
\right).
\end{equation}

{\em Property 6.}
\begin{equation}\label{p6}
z^{\rho} H_{p,q}^{m,n}\left( z \; \vline
\begin{array}{c}
(a_j,A_j)_{j=1, \ldots ,p}\\
(b_j,B_j)_{j=1, \ldots ,q}
\end{array}
\right) = H_{p,q}^{m,n}\left( z \; \vline
\begin{array}{c}
(a_j+\rho A_j,A_j)_{j=1, \ldots ,p}\\
(b_j+\rho B_j,B_j)_{j=1, \ldots ,q}
\end{array}
\right).
\end{equation}

\newpage

\section*{Figure Legends}

\begin{description}

\item{\bf Figure 1:} A log-log plot of the mean squared
displacement obtained by
numerically simulating the underlying CTRW process for
a Levy type anomalous diffusion with $\gamma=0.5$.

\item{\bf Figure 2:} a) Comparison of the theoretical FPT
distribution (solid line) with the distribution (dashed line)
obtained by numerically simulating the underlying CTRW process for
a Levy type anomalous diffusion  (with $\gamma=0.5$) that is
not close to the generalized diffusion limit ($\tau = 10^{-2}$).
b) Same comparison as above but closer to the generalized
diffusion limit ($\tau = 10^{-4}$).

\item{\bf Figure 3:} Plot of Kullback-Leibler information
criterion against $1/\tau$.

\end{description}

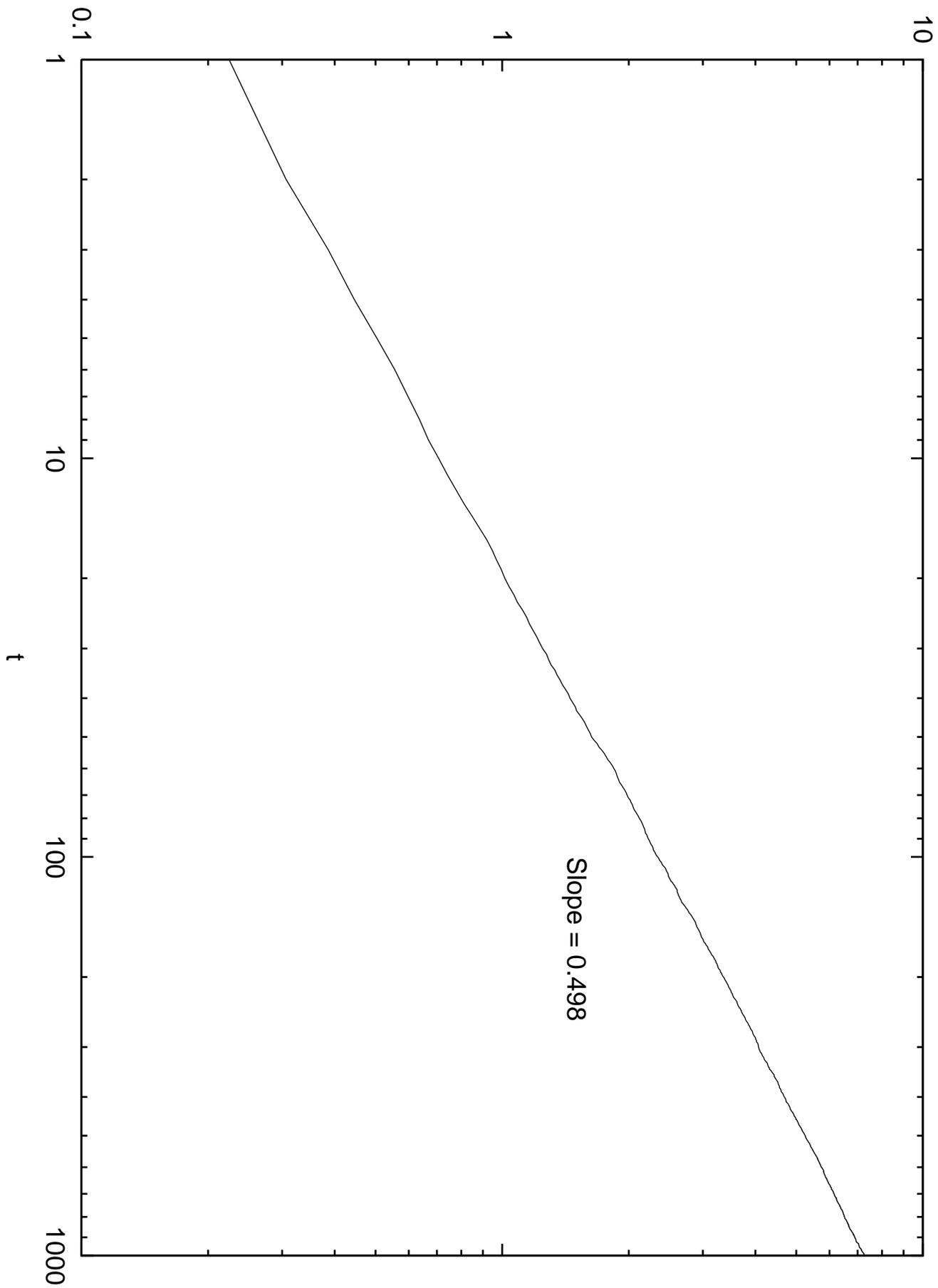
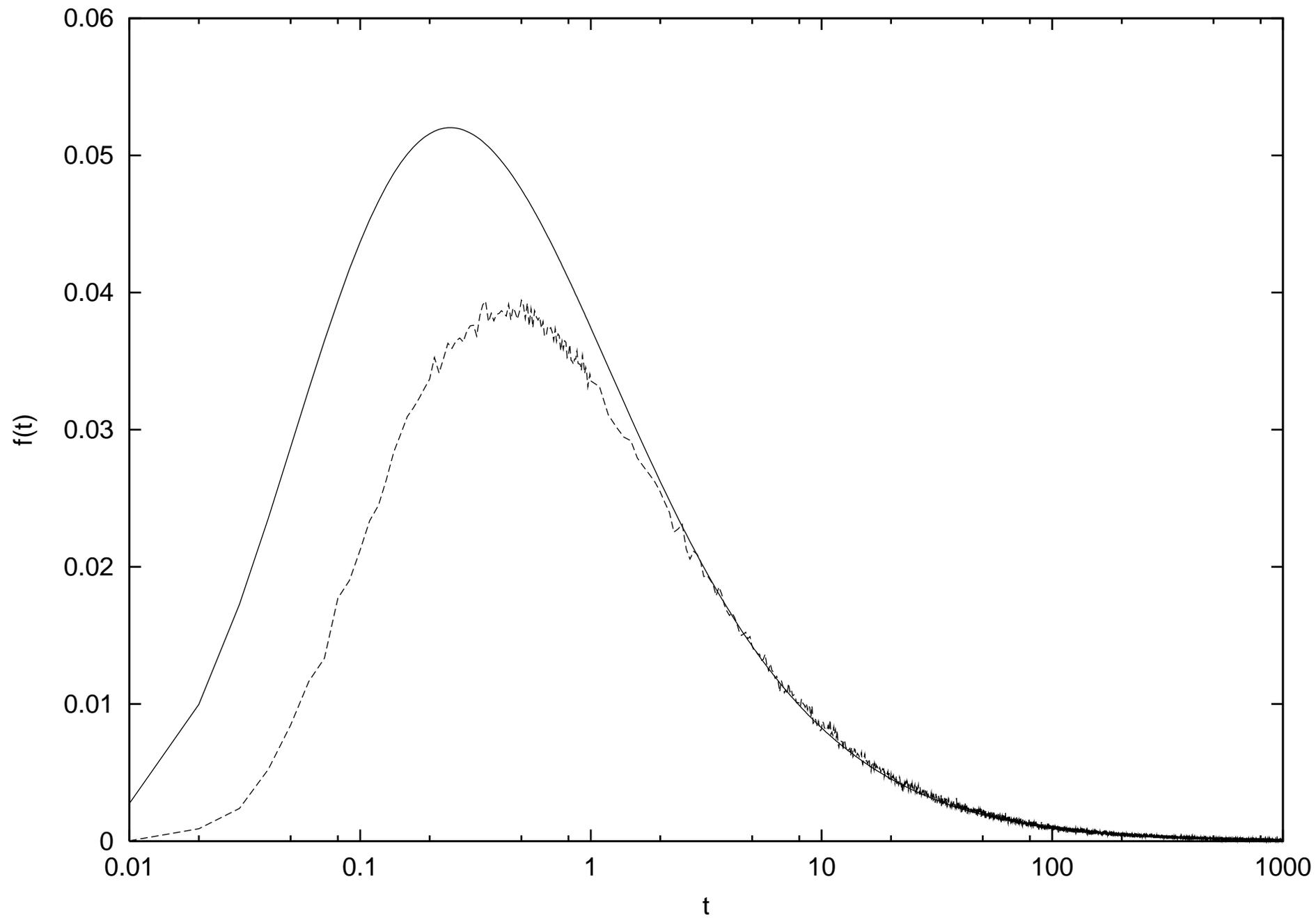
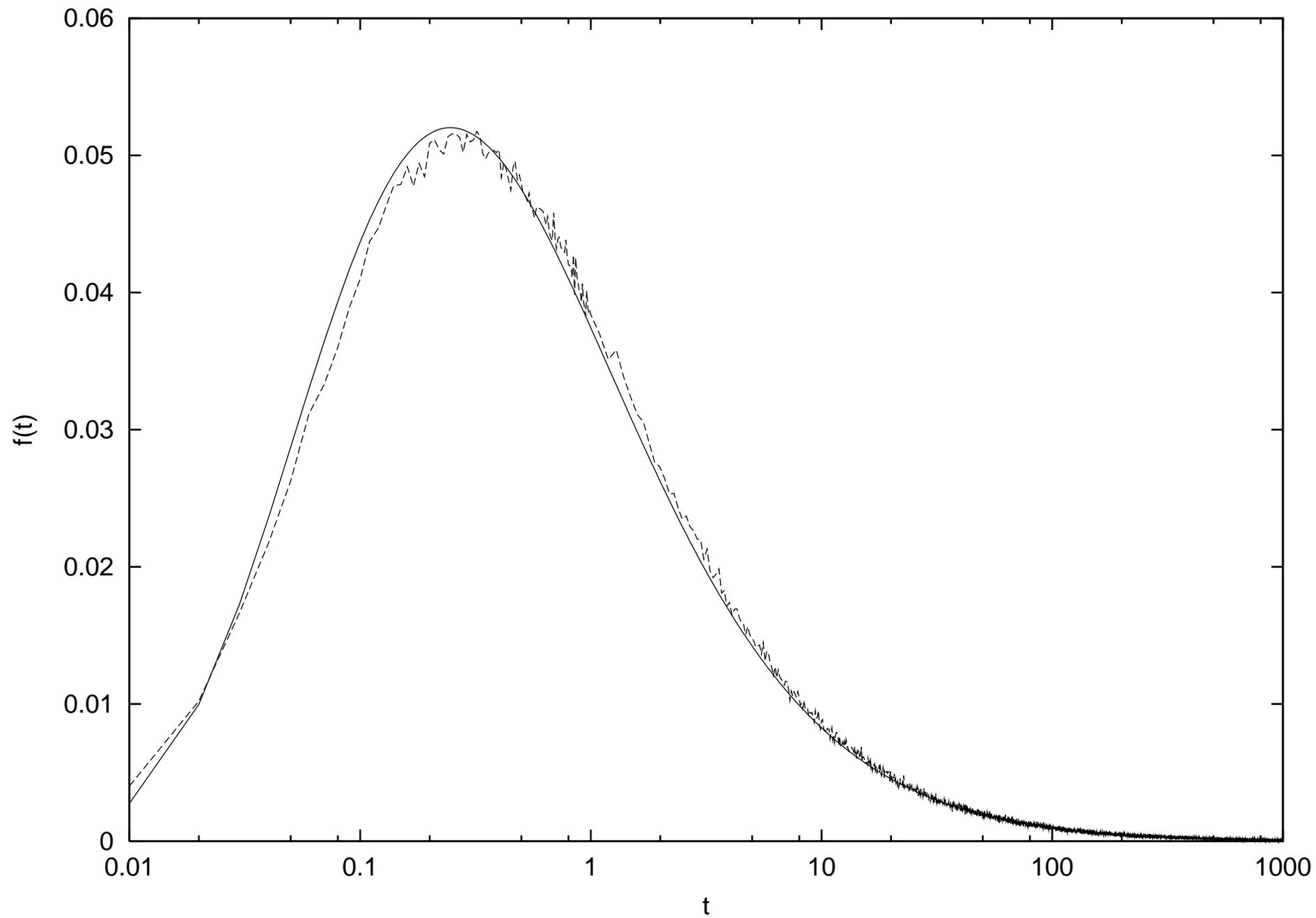
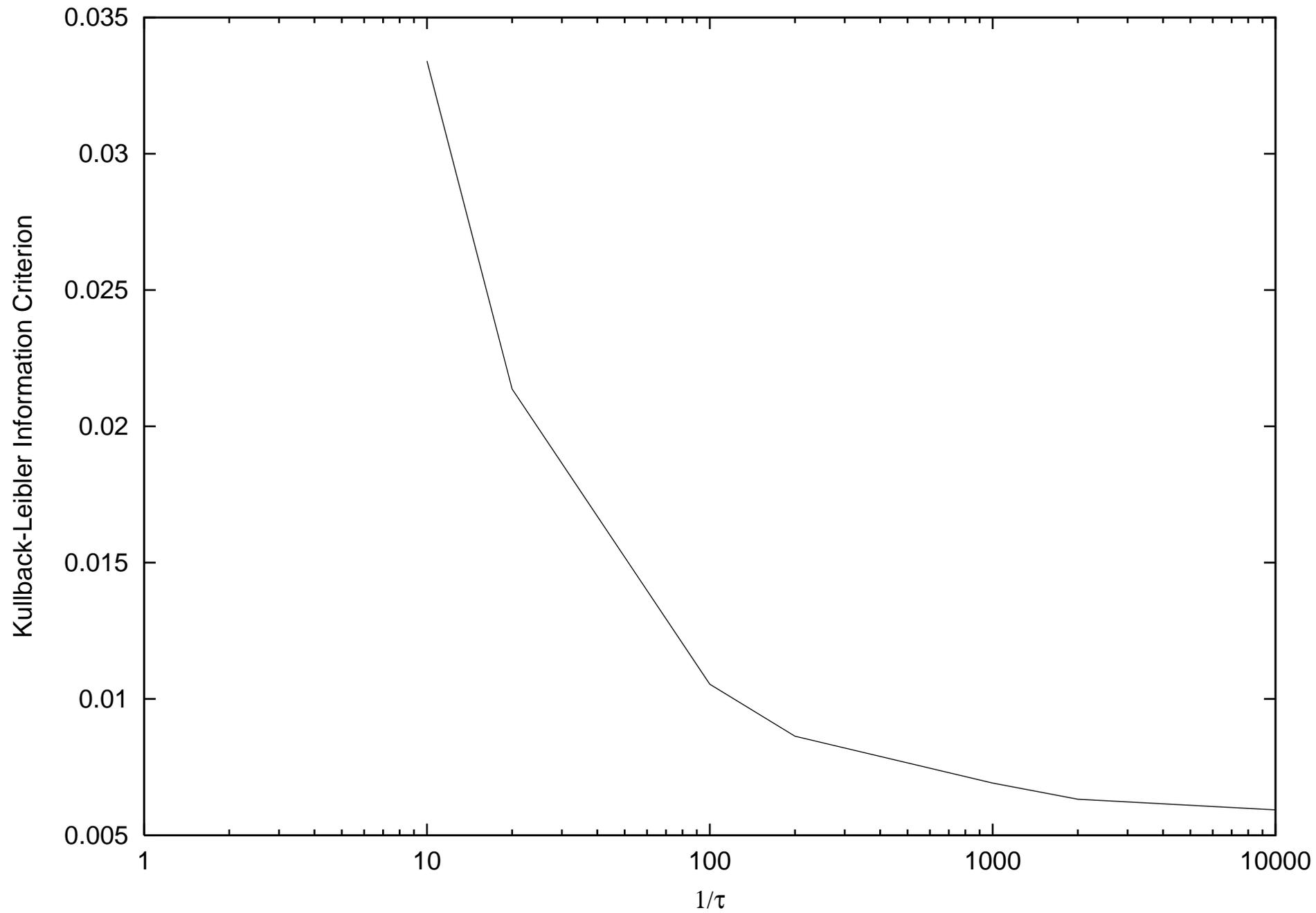
\end{document}